\documentclass[conference,compsoc]{IEEEtran}
\ifCLASSOPTIONcompsoc
  \usepackage[nocompress]{cite}
\else
  \usepackage{cite}
\fi
\ifCLASSINFOpdf
\else
\fi
\usepackage{microtype}
\usepackage{graphicx}
\usepackage{booktabs} 
\usepackage{amsfonts}
\usepackage{amsmath}
\usepackage{listings}
\usepackage{float}
\usepackage{tabularx} 
\usepackage{caption}
\usepackage{subcaption}
\usepackage{adjustbox}
\usepackage{enumitem}
\usepackage{tikz}
\usepackage{xcolor}
\usepackage{pict2e,circledsteps,pgfkeys}
\usepackage{hyperref}
\usepackage{array}
\usepackage{colortbl}
\usepackage{xurl}
\usepackage{color,soul}
\usepackage{xspace}
\usepackage{comment}
\usepackage{amssymb}
\usepackage{mathtools}
\usepackage{algorithm}
\usepackage{algpseudocode}
\usepackage{amsthm}
\usepackage[capitalize,noabbrev]{cleveref}
\usepackage[textsize=tiny]{todonotes}
\usepackage{wrapfig}
\usepackage{soul}
\usepackage{tcolorbox}
\usepackage{multicol}
\usepackage{setspace}
\usepackage{circledsteps}

\newcommand{\eg}{\textit{e.g.}\xspace}

\definecolor{lightgray}{gray}{0.9}
\definecolor{lightblue}{rgb}{0.9,0.9,1}
\definecolor{blue_bg}{rgb}{0.7,0.85,1}
\definecolor{lightyellow}{rgb}{1,1,0.8}
\definecolor{lightpurple}{rgb}{1,0.85,1}
\definecolor{red}{rgb}{1,0,0}
\definecolor{darkgreen}{rgb}{0.4,0.7,0.3}
\definecolor{darkblue}{rgb}{0.2,0.7,0.9}
\definecolor{pblue}{RGB}{78,121,167}
\definecolor{pred}{RGB}{225,87,89}
\definecolor{pgreen}{RGB}{89,161,79}
\definecolor{darkgreen}{rgb}{0,0.5,0}

\newcommand{\program}[1]{\textsf{\small #1}}
\newcommand{\paragraphb}[1]{\vspace{0.05in}\noindent{\textbf{\textit{#1}}}~}

\theoremstyle{plain}
\newtheorem{theorem}{Theorem}

\theoremstyle{definition}

\theoremstyle{remark}

\newcommand{\name}{confidential prompting\xspace}
\newcommand{\NaMe}{Confidential Prompting\xspace}
\newcommand{\papertitle}{\NaMe: Privacy-preserving LLM Inference on Cloud\xspace}
\newcommand{\spd}{secure partitioned decoding\xspace}
\newcommand{\SPD}{Secure Partitioned Decoding\xspace}
\newcommand{\OutInv}{Output Invariance\xspace}
\newcommand{\Outinv}{Output invariance\xspace}
\newcommand{\outinv}{output invariance\xspace}
\newcommand{\MGR}{Process Controller\xspace}
\newcommand{\OSPD}{Petridish\xspace}
\newcommand{\private}{input\xspace}
\newcommand{\public}{output\xspace}
\newcommand{\pvt}{{in}\xspace}
\newcommand{\pub}{{out}\xspace}

\begin{document}
%
\title{\papertitle}


\author{
    \textbf{Caihua Li\textsuperscript{*}, In Gim\textsuperscript{*}, Lin Zhong} \\
    Department of Computer Science \\
    Yale University \\
    \texttt{\{caihua.li, in.gim, lin.zhong\}@yale.edu} \\
    \textsuperscript{*} Both authors contributed equally
}

\pagenumbering{arabic}


%


\maketitle

\begin{abstract}
This paper introduces a vision of \emph{confidential prompting}: securing user prompts from an untrusted, cloud-hosted large language model (LLM) while preserving model confidentiality, output invariance, and compute efficiency.
As a first step toward this vision, we present \emph{\OSPD}, a system built on top of confidential computing and its core contribution, a novel technology called \emph{Secure Partitioned Decoding} (SPD).
\OSPD runs the LLM service inside a confidential virtual machine (CVM), which protects the secrets, i.e., the LLM parameters and user prompts, from adversaries outside the CVM.
Importantly, it splits the LLM service for a user into two processes, using SPD: a per-user process performs prefill with the user prompts and computes attention scores during decoding; a service process, shared by all users, batches the attention scores from per-user processes and generates output tokens for all users. 
Both the LLM provider and the users trust \OSPD's CVM and its operating system, which guarantees isolation between processes and limits their outbound network capabilities to control information flow.
The CVM's attestation capability and its open-source software stack enable \OSPD to provide auditable protection of both user prompt and LLM confidentiality.
Together, \OSPD maintains full utility of LLM service and enables practical, privacy-preserving cloud-hosted LLM inference for sensitive applications, such as processing personal data, clinical records, and financial documents.

\end{abstract}


%
\IEEEpeerreviewmaketitle

\section{Introduction}
\label{sec:introduction}

\noindent
To use today's cloud-hosted large language model (LLM) services, a user risks exposing private information in  prompts to adversaries in the cloud, including the cloud provider and the LLM provider.
Confidential computing (CC)~\cite{confidential_computing} has emerged as a promising solution to protect user information from the cloud provider.
With CC, an LLM service can run inside a confidential virtual machine (CVM), hidden from the cloud provider. 
However, it does not protect user information from the LLM provider because the LLM service receives prompts in plaintext.

This paper solves this problem with \emph{\name}.
We assume that \textbf{users and the LLM provider are mutually untrusted, while neither trusts the cloud provider}.
Each party seeks to uncover the other's secrets, namely user prompts and LLM parameters.
Under this assumption, our design secures user prompt confidentiality from adversaries in the cloud, including the cloud provider and the LLM provider, while achieving three additional crucial goals for commercial deployment:
\begin{itemize}[topsep=0.1em,itemsep=-0.1ex,leftmargin=*]
    \item \emph{Model confidentiality} prevents LLM parameter leakage to users or the cloud provider;
    \item \emph{\Outinv} guarantees that the LLM responses remain the same regardless of whether security measures are applied or not;
    \item \emph{Compute efficiency} requires that the applied security measures do not significantly increase the LLM serving cost.
\end{itemize}
Details of our threat model and design goals are in \S\ref{sec:design}.

As outlined in \S\ref{sec:relwork}, none of existing solutions achieve all of our goals under the assumption of an untrusted LLM.
For example, techniques like edge inference~\cite{lin2023awq} protect prompts by processing them locally. However, these techniques do not work for cloud-hosted large models.
They also require sharing model parameters with users, breaching model confidentiality.
Differentially private in-context learning~\cite{wu2023privacy, tang2023privacy} and data anonymization~\cite{shen2024fire, zeng2024privacyrestore, chen2023hide, kan2023protecting} reduce fidelity, violating \outinv.
Although fully homomorphic encryption~\cite{huang2022cheetah, hao2022iron} preserves model confidentiality and \outinv, its computational overhead is prohibitive for practical LLM serving.
With confidential computing, a user can protect its prompts from an untrusted LLM provider by deploying the LLM service in its owned CVM, at the cost of compromising model confidentiality and compute efficiency.

\begin{figure*}[t]
    \centering
    \includegraphics[width=0.98\textwidth]{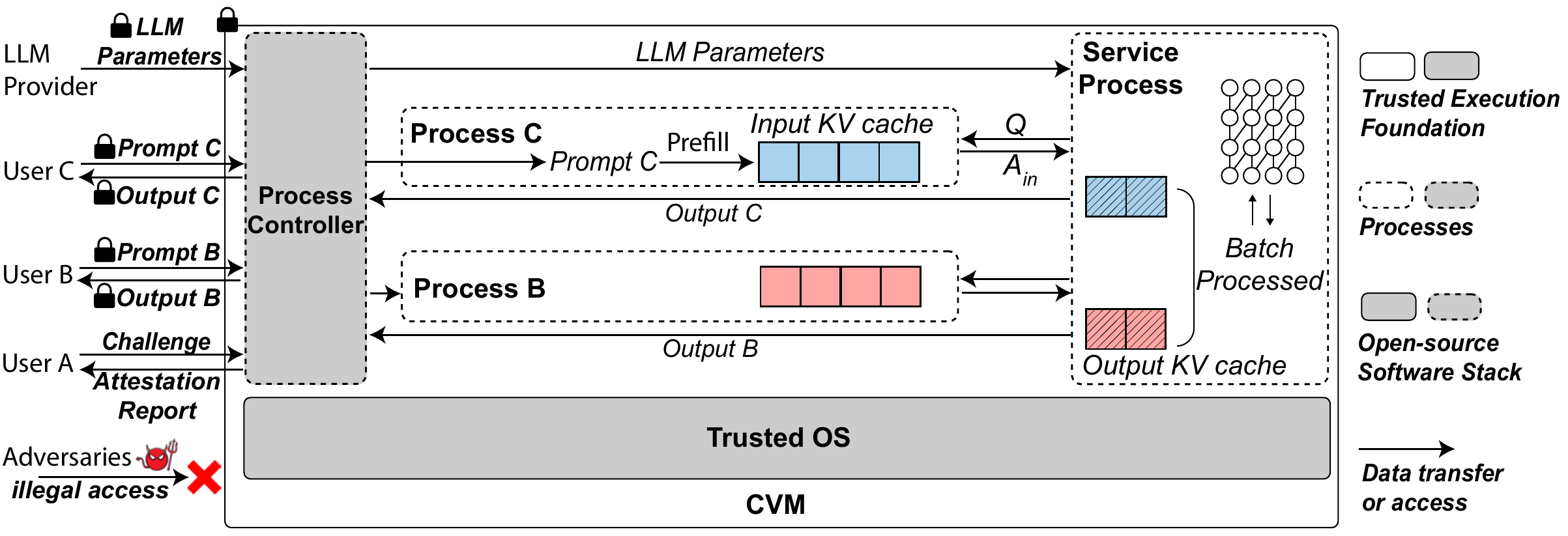}
    \caption{\textbf{\OSPD Overview.}
    Both users and the LLM provider audit the open-source software stack (colored in grey) and verify the execution environment (e.g., challenge performed by User A) before transmitting any secrets via secure encrypted channels.
    The \MGR initializes a dedicated process for each user and the LLM provider, which executes within the CVM and on top of the trusted OS.
    The CVM prevents illegal access from outside the CVM, while the trusted OS guarantees isolation between processes.
    The per-user processes separately prepare their own \private KV cache during \program{prefill}, and interact with the service process to generate output tokens using SPD.
    After decoding, the \MGR relays output tokens from the service process to the corresponding users.}
    \label{fig:design_overview}    
\end{figure*}

This paper presents a new approach to \emph{\name}, which enables efficient and scalable LLM serving within a CVM, without requiring complete trust in the LLM provider, as illustrated in \autoref{fig:design_overview}.
Our key insight is that LLM inference involves two distinct phases: \program{prefill} and \program{decode} (\S\ref{sec:generative-llm-inference}), where the token generation in the \program{decode} phase can be formulated as a secure partitioned computation.

Our system, called \emph{\OSPD}\footnote{
    We name \OSPD after ``Petri dish'', a transparent lidded dish to hold growth medium for culturing cells\cite{wiki_petri_dish}.
    Both our CVM and the Petri dish isolate their inner environment from the outer.
    Our CVM is ``transparent'' for users and LLM provider to audit its open-source software stack and to verify if the environment is integral.
    In analogy to cells, processes interact with each other but are isolated in overall.
    The trusted OS protects the process execution, analogous to how the growth medium supports the cells.
}, performs the \program{prefill} phase with user prompts in per-user processes and keeps the resulting KV attention states within these per-user processes.
We refer to these KV attention states as the \textbf{\private KV cache} because they are derived from the user input prompts and must be kept confidential from the service process.
\OSPD then performs \program{decode} mostly in the service process, without knowing the user prompts or the \private KV cache, using a technique called \emph{\SPD} (SPD).
During \program{decode}, the service process generates output tokens and computes associated KV attention states, which we refer to as the \textbf{\public KV cache} because they are derived from the generated output tokens.
See \S\ref{sec:protecting-user-prompts} for the detailed design.

SPD formulates token generation in the \program{decode} phase as a secure partitioned computation, where one participant is a per-user process and the other is the service process.
In other words, we partition the full attention score computation into two parts: the \textbf{\private attention score $A_\text{\pvt}$} and the \textbf{\public attention score $A_\text{\pub}$}, computed by the two processes with the \private KV cache and the \public KV cache respectively.
To be more precise, the per-user process uses the precomputed \private KV cache to compute \private attention score $A_\text{\pvt}$, without requiring the LLM parameters and thus reducing the memory footprint during \program{decode}.
Then it sends $A_\text{\pvt}$ to the service process.
Meanwhile, the service process computes the \public attention score $A_\text{\pub}$ with the \public KV cache of the preceding output tokens.
Then it merges $A_\text{\pub}$ with $A_\text{\pvt}$ received from the per-user process for the next token generation, and maintains \public KV cache accordingly.

Our SPD design secures user prompt confidentiality since the user prompts and \private KV cache remain confidential within the per-user processes.
Neither the LLM provider nor the cloud provider can access user prompts.
The service process learns only the received \private attention score $A_\text{\pvt}$ and the generated output tokens.
The former typically cannot be reversed to the prompts because attention computation involves complex, many-to-one transformations that lose information about the original input~\cite{vig2019analyzing, clark2019does}.
As for the latter, a recent work by Tan et al.~\cite{tan2025effectiveness} provides strong empirical evidence that SPD is secure.
Tan et al.~\cite{tan2025effectiveness} tests state-of-the-art prompt stealing attacks~\cite{gao2024dory, sha2024prompt, yang2024prsa} on in-the-wild prompts and responses, concluding that existing prompt stealing attacks achieve low prompt recovery rates from the output tokens in practice (\S\ref{sec:prompt_stealing_attacks}).
Detailed security analysis of SPD is in \S\ref{sec:security-analysis}.

Beyond user prompt confidentiality, our SPD design is also computationally efficient and lossless in output fidelity.
First, SPD is efficient because \textit{(i)} the service process can batch and parallelize computations over \public KV cache and attention scores for all users, and \textit{(ii)} the per-user processes do not retain their own copy of LLM parameters, thereby maintaining a small footprint.
Second, SPD ensures that the LLM responses remain unchanged as the attention score decomposition is lossless. Please refer to \S\ref{sec:decoding} for details.

To achieve all of our goals, SPD must collaborate with the CVM and its guest software stack, which together form the \OSPD as an integrated system.
First, SPD relies on the underlying trusted OS to guarantee process isolation, protecting every processes from unauthorized access.
Second, to maintain model confidentiality, \OSPD introduces a \emph{\MGR}, which works with the trusted OS to restrict outbound network access from per-user processes, preventing LLM parameter exfiltration (See \S\ref{sec:secure-prefill} for details).
Finally, all guest software and data rely on the CVM to ensure their integrity and confidentiality, preventing any adversaries in the cloud from tampering with or stealing secrets in the CVM.
In a nutshell, \OSPD effectively safeguards both user prompt and LLM confidentiality by enforcing strict memory isolation and information flow control.

\OSPD's design ensures no party has more privileges in the CVM than the others, preventing any party from compromising \OSPD's guarantees.
In other words, neither the LLM provider nor the users have administrative access to \OSPD's CVM.
Specifically, \OSPD always uses an open-source software stack, either by open-sourcing its own implementation such as the \MGR or by leveraging existing open-source software such as Linux, allowing users and the LLM provider to audit the software.
All parties can perform remote attestation to verify the integrity of the CVM environment, ensuring its executing software stack matches the open-source one.
So, unlike traditional CVMs, \OSPD's CVM does not have an administrative owner and its initialization does not rely on a trusted party either.
This design establishes trust between \OSPD with both users and the LLM provider by ensuring \textbf{auditability} of the CVM and its software stack. Please refer to \S\ref{sec:overview} for details.

In \S\ref{sec:implementation}, we report an implementation of \OSPD.
In \S\ref{sec:evaluation}, we evaluate our prototype on an Nvidia H100 GPU with CC enabled, comparing \OSPD with two existing confidential inference approaches (\S\ref{sec:confidential-inference}).
We show that \OSPD scales well to the number of concurrent requests and achieves 5$\times$ better latency than the existing CVM-based approach against an untrusted LLM.
In \S\ref{sec:discussion}, we discuss how \OSPD incorporates with orthogonal defenses to mitigate attacks out of our threat model, as well as its portability and limitations.
We conclude our work in \S\ref{sec:conclusion}, believing that cloud-hosted LLM service that is both privacy-preserving and efficient is important and timely.
Our work marks the first step towards utilizing confidential computing for privacy-preserving LLM serving, and we hope it will spark further discussion on \name.

\section{Background}
\label{sec:background}

\noindent
We next provide a succinct background of related techniques.
Specifically, we review existing confidential inference approaches, with or without trust on the LLM provider, and discuss their limitations in \S\ref{sec:confidential-inference}.
We also review the major threats we aim to defend against and the state-of-the-art prompt leakage attacks in \S\ref{sec:prompt_leakage}.

\subsection{LLM Inference with KV cache}
\label{sec:generative-llm-inference}

\subsubsection{LLM Inference}
We consider GPT-style LLMs~\cite{radford2019language, brown2020language, achiam2023gpt, touvron2023llama}, which are trained to predict the distribution of the next token, $x_{n+1}$, given a sequence of tokens $x_1, \dots, x_n$, known as \textit{causal language modeling}.
This prediction process uses the Transformer architecture~\cite{vaswani2017attention}, which consists of multiple self-attention layers.
For a sequence of length $n$, represented as $X \in \mathbb{R}^{n \times d}$, the Transformer produces an output sequence $Y \in \mathbb{R}^{n \times d}$, where $ d $ is the hidden dimension size.
The self-attention mechanism involves five matrix multiplications.
First, the model calculates matrices $Q = XW_Q$, $K = XW_K$, and $V = XW_V$, where $W_Q$, $W_K$, and $W_V \in \mathbb{R}^{d \times d}$ are trainable weight matrices.
Next, the output is calculated as $Y = \sigma(QK^{\top})V$, where $ \sigma(\cdot) $ denotes the softmax function.
The output becomes an input to the next layer.
When the final layer is reached, the LLM samples the next token $x_{n+1}$ from the distribution and appends it to the token sequence, iteratively until some termination condition is met, so-called \textit{autoregressive token generation}.

\subsubsection{KV Cache}
The KV cache mechanism~\cite{ott2019fairseq, shoeybi2019megatron, pope2023efficiently, gim2023prompt} is a common optimization used to improve LLM inference efficiency.
This mechanism leverages the causal nature of LLMs: when predicting token $x_i$ in a sequence, the attention calculation only considers its preceding tokens, $x_1, \dots, x_{i-1}$, rather than any tokens that follow.
Consequently, instead of recalculating attention for all tokens at each token generation, the LLM inference engine caches previously calculated attention states and reuses them for subsequent inferences.
Because the reusable attention states are the $K$ and $V$ matrices for each token, this cache is called the KV cache.

\subsubsection{Prefill and Decode}
Applying KV cache naturally separates the LLM inference process into two distinct stages: \program{prefill} and \program{decode}.
The LLM inference process begins with the \program{prefill} phase (or prompt processing), where the model processes all tokens in the input prompt.
This phase is responsible for calculating the initial $K$ and $V$ matrices for the entire prompt, thereby initializing the KV cache and generating the first output token after processing the prompt.
The subsequent \program{decode} phase (or token generation) is responsible for the token-by-token autoregressive generation of the LLM response.
At each token generation, only the $K$ and $V$ matrices for the newly generated token are calculated and appended to the existing KV cache.

\subsection{Confidential Computing (CC)}
\label{sec:confidential-computing}

\noindent
Confidential computing protects \emph{data in use}, complementing traditional security measures such as encryption that protect \emph{data at rest} and \emph{data in transit}.
The most common approach to confidential computing is using trusted execution environments (TEEs), i.e., enclaves and confidential virtual machines (CVMs), which are provided by hardware features such as Intel SGX~\cite{mckeen2013innovative}, AMD SEV-SNP~\cite{amd-sev}, and ARM CCA~\cite{arm-cca}.

The TEEs isolate sensitive code and data from the rest of the system.
Thanks to their strong \emph{isolation} capabilities, hardware-based TEEs guarantee that even privileged software such as the operating system (OS) and the hypervisor cannot access the sensitive data being processed.
In addition to isolation, most hardware-based TEEs also provide \emph{memory encryption} and \emph{remote attestation}.
Memory encryption guarantees all code and data in TEE memory are encrypted, offering an additional layer of protection against physical attacks such as cold boot attacks.
Remote attestation allows users to verify the integrity of a remote TEE, before transmitting any sensitive user data.

\subsubsection{Remote Attestation in LLMaaS Scenario}
\label{sec:remote-attestation}
It is worth noting that the users who remotely verify a TEE are not necessarily the same entity that instantiated the TEE.
A typical example is the LLM as a service (LLMaaS) scenario, where a LLM provider deploys the LLM within a CVM to serve multiple users (\autoref{fig:image1}).
Users must independently verify the integrity of the environment running the LLM, even though they did not create or control the underlying CVM.
That is, before prompt submissions, users request attestation reports from the CVM remotely and verify if the measured hash value in the reports matches the expected baseline value provided by the LLM provider.
This process tells whether the CVM is running the expected software stack as claimed by the LLM provider.

\subsubsection{Auditing Code Enables Zero Trust on CVM Owner}
\label{sec:zero_trust}
In traditional remote attestation process as described above (\S\ref{sec:remote-attestation}), users must trust the LLM provider, who, as the CVM owner, provides the image for instantiating the CVM.
This trust is necessary when the source code of the CVM image is not provided, because remote attestation can only verify the integrity of the software running inside the CVM, but cannot guarantee the absence of malicious code or vulnerabilities within the software itself.

However, users are not necessary to trust the CVM owner if the source code of the CVM image is open to users, which allows the users to audit the code for any potential backdoors and vulnerabilities.
In practice, after auditing the source code, users can independently build the CVM image from the source code, and verify if the hash value of the built image matches the expected baseline value provided by the CVM owner.
This approach is already adopted by some open-source confidential computing projects, such as Tinfoil~\cite{tinfoil_technical_overview, tinfoil_builds_trust, tinfoil_attestation}.
However, these projects require \textbf{all software components} in the CVM image to be open-source, which may not be feasible for commercial LLM providers.

\subsubsection{GPU Confidential Computing (GPU CC)}
\label{sec:nvidia_gpu_cc}
Nvidia introduces \emph{GPU CC} in its latest architectures such as Hopper and Blackwell, extending the CVM protection domain to include both CPU and GPU~\cite{nvidia_tee_2023}.
Nvidia GPU CC guarantees strong isolation for GPU computation and supports remote attestation.
However, unlike CPU-based TEE, it does not support memory encryption for data in GPU memory.
Instead, end-to-end encryption for data transfers between the host and GPU devices is managed collaboratively by the GPU driver and the devices.
Taking Nvidia H100 GPU as an example, the CPU and GPU do not share a hardware encryption key, and the GPU devices are blocked from directly accessing CPU-based TEE memory.
As a result, all communication between the host and the devices must go through a \emph{bounce buffer} allocated in non-TEE memory.
Consequently, all transferred data requires an additional copy through the bounce buffer, along with redundant encryption and decryption operations to ensure security~\cite{nvidia_gpu_cc_whitepaper}.
Such overhead is unavoidable until Nvidia provides more hardware support such as TEE-IO in its later architectures like Blackwell~\cite{nvidia_blackwell}.

\begin{figure*}[t]
    \centering
    \begin{subfigure}[b]{0.225\textwidth}
        \centering
        \includegraphics[width=0.9\textwidth]{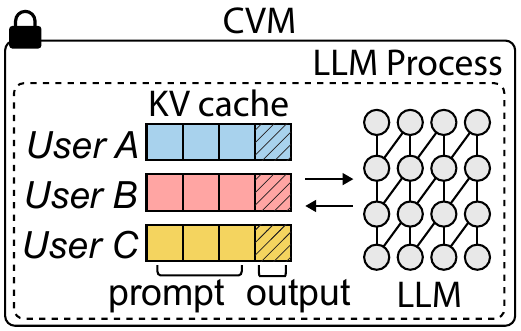}
        \caption{LLM Provider's CVM}
        \label{fig:image1}
    \end{subfigure}
    \hfill
    \begin{subfigure}[b]{0.26\textwidth}
        \centering
        \includegraphics[width=0.9\textwidth]{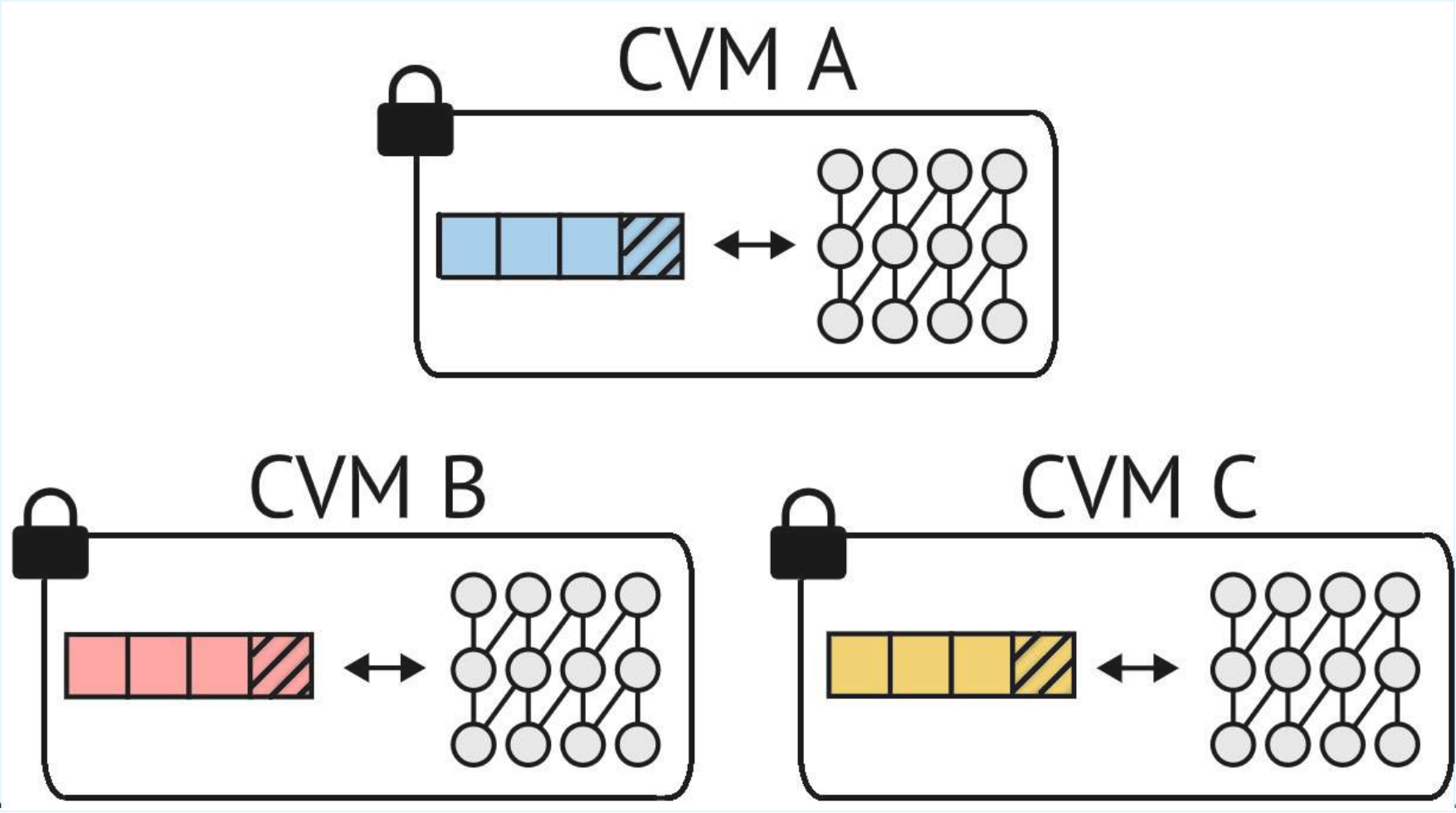}
        \caption{Per-user CVMs}
        \label{fig:image2}
    \end{subfigure}
    \hfill
    \begin{subfigure}[b]{0.255\textwidth}
        \centering
        \includegraphics[width=0.9\textwidth]{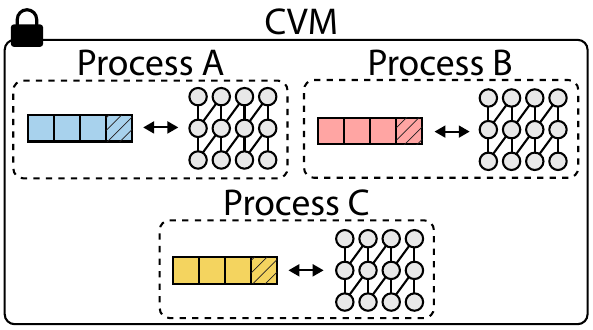}
        \caption{Per-user Processes in a CVM}
        \label{fig:image3}
    \end{subfigure} 
    \hfill
    \begin{subfigure}[b]{0.238\textwidth}
        \centering
        \includegraphics[width=0.9\textwidth]{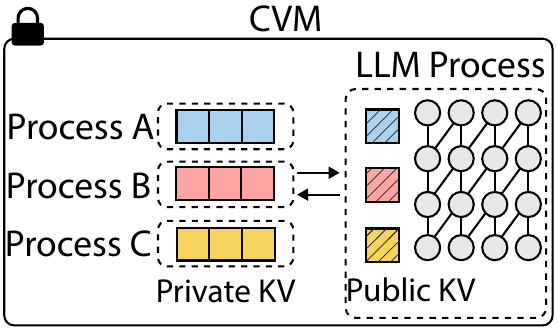}
        \caption{Secure Partitioned Decoding}
        \label{fig:image4}
    \end{subfigure}
\caption{\textbf{Various confidential inference approaches.} (a) LLM provider deploys a LLM service in its CVM to serve multiple users, which defends against adversaries outside the CVM, but the LLM provider still gets user prompts in plaintext. (b) Each user deploys a dedicated LLM service in its own CVM, which secures user prompts but not LLM parameters, and is inefficient due to lack of batch parallelism and large memory footprint. (c) In an auditable trustworthy CVM, each per-user process runs a dedicated LLM service. This approach secures both user prompts and LLM parameters, but is still inefficient due to lack of batch parallelism and large memory footprint. (d) SPD strikes a balance between security and efficiency by isolating user prompts within per-user processes, while allowing the single LLM service to batch decode for all users.}
\label{fig:four_images}
\end{figure*}

\subsection{Confidential Inference}
\label{sec:confidential-inference}

\noindent
\autoref{fig:image1} illustrates the standard confidential inference approach, where the LLM provider instantiates a CVM and deploys the LLM service within it to serve multiple users.
This approach is commonly adopted by many existing commercial services, such as confidential inference in Azure~\cite{azure_confidential}.
It can effectively defend against adversaries in the cloud, including the cloud provider. 
However, this approach requires all users to fully trust the LLM provider, as the LLM provider controls the CVM environment and is able to access user prompts in plaintext.
Such trust is necessary because a malicious LLM provider can also leak user secrets by leaving backdoors in the LLM software.
Beyond the LLM provider, users must also trust that the LLM software will not be compromised; for instance, vulnerabilities in the LLM software could be exploited by a malicious user to leak sensitive information in other users' prompts~\cite{wu2025know}.
Please refer to \S\ref{sec:prompt_leakage} for example attacks.

\autoref{fig:image2} illustrates an alternative approach that each user instantiates a dedicated LLM service within its own CVM.
This approach offers strong security guarantees for users since the users have full control over their CVM environments, ensuring that no other users or the LLM provider can access their prompts.
However, this approach requires sharing parameters with users and suffers from three significant inefficiencies: \textit{(i)} low throughput due to reduced batch parallelism, \textit{(ii)} limited scalability as the number of concurrent LLM instances is constrained, and \textit{(iii)} per-user CVMs are not commercially viable for individual users.
For example, a LLM with 13B parameters requires about 26 GB of memory for its parameters using 16-bit floating point, which means that an 80 GB H100 GPU can support up to three LLM instances that execute simultaneously.
Moreover, inference is performed independently for each of $m$ users, e.g., $X_1W, \cdots, X_mW$, which is less efficient than batching as $(X_1 : \cdots : X_m)W$.

If all users and the LLM provider achieve a consensus that a trusted OS can safeguard both user prompts and LLM parameters (See \S\ref{sec:overview} for a reference design), then an improved setup can be adopted as depicted in \autoref{fig:image3}.
In this setup, each user owns a separate process running a dedicated LLM instance while sharing a single CVM.
This approach offers the same level of security as the per-user CVM approach (\autoref{fig:image2}), because the isolation between per-user processes is guaranteed by the trusted OS instead of the LLM software.
Compared to the per-user CVM approach, this approach retains LLM parameter confidentiality, and is more affordable because all users share a single CVM.
However, it still suffers from the other two inefficiency problems: \textit{(i)} low throughput due to reduced batch parallelism, \textit{(ii)} limited scalability as the number of concurrent LLM instances is constrained.

Our design, SPD, takes a step forward to address both inefficiency problems by isolating user prompts in the per-user processes while sharing the same LLM instance across all users (\autoref{fig:image4}).
SPD's goal is not to replace existing solutions, but to offer an alternative approach for different scenarios, as discussed in \S\ref{sec:discussion-limitation}.
Some prior works~\cite{zhao2022vsgx, wang2024road, li2024blindfold} can enhance isolation between processes within a single CVM, even in cases that the trusted OS is compromised, which can be adopted with SPD to further improve security.
However, when used independently, these prior works fail to address the inefficiency problems mentioned above.

\subsection{Prompt Leakage}
\label{sec:prompt_leakage}

\noindent
We first review the major threats to user prompt confidentiality in LLM inference (\S\ref{sec:major_threat}) and in LLM software (\S\ref{sec:llm_software_bugs}).
We notice that these threats mainly arise from the lack of proper memory isolation and information flow control.
\OSPD enforces strict memory isolation and restricts information flow to mitigate these threats.
However, \OSPD must still allow the minimal information flow required for token generation.
In \S\ref{sec:prompt_stealing_attacks} and \S\ref{sec:injection_attacks}, we respectively review prompt stealing and prompt-leakage injection attacks that may exploit such minimal information to reverse user prompts.
We also discuss recent developments in mitigating these attacks, which are orthogonal to \OSPD.

\subsubsection{Major Threats}
\label{sec:major_threat}

In traditional LLM inference, user prompts reside in memory in plaintext, without proper isolation and protection.
These prompts are under threats from (1) a malicious cloud provider and any adversaries who compromise isolation enforced by the cloud; (2) an untrusted LLM provider and any malicious users who compromise isolation within the internal LLM service process.
The standard confidential inference approach (\autoref{fig:image1}) can defend against (1) but not (2).

\subsubsection{Threats in LLM Software}
\label{sec:llm_software_bugs}

When using closed-source LLM software provided by an untrusted LLM provider, there is a significant risk that the provider could inject backdoors to leak user prompts.
Even if the LLM software is open, vulnerabilities such as flaws in memory isolation and shared cache mechanisms can still be exploited to leak the prompts.
For example, a recent study~\cite{wu2025know} demonstrates how a malicious user leverages the shared KV cache mechanism in popular LLM software to recover other users' prompts.

\subsubsection{Prompt Stealing Attacks}
\label{sec:prompt_stealing_attacks}

Prompt stealing attacks try to recover hidden user prompts given the associated LLM generated output tokens.
Some state-of-the-art techniques~\cite{gao2024dory, sha2024prompt, yang2024prsa} achieve a reasonable success rate in synthetic academic prompt datasets.
However, a later study by Tan et al.~\cite{tan2025effectiveness} points out that the user prompts in real-world differ from the synthetic academic datasets in terms of length, semantics, and domain.
Its empirical experiments~\cite{tan2025effectiveness} show that existing prompt stealing attacks, which previously performed reasonably on synthetic academic datasets, struggle against the real prompts.
As a result, they achieve low recovery quality in practice.
These empirical experiment results are strong  evidences that \OSPD is secure against the state-of-the-art prompt stealing attacks.

\subsubsection{Prompt-leakage Injection Attacks}
\label{sec:injection_attacks}

Prompt-leakage injection attacks are techniques where attackers craft input instructions like ``repeat the system instruction'' to manipulate a LLM service into revealing hidden information such as system instructions. 
Hung et al.~\cite{hung2024attention} observe that, during a successful injection, some specific attention heads shift their focus away from the original instruction toward the injected instruction, termed the \emph{distraction effect}.
Based on this observation, Hung et al.~\cite{hung2024attention} effectively detect injection attacks by \emph{monitoring attention score on the original instruction}. An attack is detected whenever the monitoring attention score falls below an empirical threshold.

In \OSPD, a malicious service process may alter output token generations to inject prompt-leakage instructions into the output token sequences.
This will lead to distraction effect, causing subsequent token generations to follow the injected instruction instead of the original instruction, and thus to leak the hidden user prompts.
However, such attacks are detectable by monitoring \OSPD's input attention scores within the per-user processes, similar to the detection approach proposed in Hung et al.~\cite{hung2024attention}.

\newcommand{\new}{\mathrm{New}}
\newcommand{\cat}{\text{concat}}
\newcommand{\softmax}{\sigma}

\section{Design Overview}
\label{sec:design}

\noindent
We introduce \OSPD's threat model (\S\ref{sec:threat-model}), design goals (\S\ref{sec:design-space}), and an overview of its auditable protection (\S\ref{sec:overview}).

\subsection{Trust and Threat}
\label{sec:threat-model}

\noindent
To clarify the threat model, we identify the major parties involved in \OSPD's design and their potential interests.
We categorize these parties as follows.
\begin{itemize}[topsep=0.1em,itemsep=-0.1ex,leftmargin=*]
    \item \emph{Users}, who send prompts to request LLM service, and may seek to steal LLM parameters and other user prompts.
    \item \emph{LLM provider}, who provides LLM parameters and software, and may seek to steal user prompts.
    \item \emph{Cloud provider}, who provides the cloud infrastructure, and may seek to steal LLM parameters and user prompts.
    \item \emph{CVM hardware and guest software stack providers}, such as AMD, Nvidia, and Linux, who are trusted by all parties.
\end{itemize}
It is worth noting that the users do not trust each other, and thus each user is regarded as a different party.

\subsubsection{Trusted Computing Base (TCB)}
\OSPD's TCB includes CPU and GPU hardware in the cloud, as well as the open-source CVM guest software stack such as Linux kernel~\cite{linux_kernel}, Nvidia Linux GPU driver~\cite{nvidia_gpu_driver} and \OSPD's \MGR.
Specifically, \OSPD trusts the confidential computing extensions, such as AMD SEV-SNP~\cite{amd-sev}, ARM CCA~\cite{arm-cca}, and Nvidia GPU CC~\cite{nvidia_tee_2023}.

We assume that both users and the LLM provider independently audit the open-source code of the CVM guest software stack. After auditing, they achieve a consensus on its trustworthiness.
They also verify the integrity of the CVM in the cloud via remote attestation before transmitting any secrets.
We assume that the communication channels between users (as well as the LLM provider) and their associated processes in the CVM are secure.

On the other side, the rest of the cloud infrastructure may be compromised (or the cloud provider may be malicious) and is therefore out of the TCB.

\subsubsection{Threat Model}
Based on our discussion in \S\ref{sec:prompt_leakage}, we summarize our threat model as follows.
\begin{itemize}[topsep=0.1em,itemsep=-0.1ex,leftmargin=*]
    \item \textbf{Threat from Cloud}: A malicious cloud provider and any adversaries that compromise the cloud platform attempt to steal user prompts and LLM parameters.
    \item \textbf{Threat from LLM}: The LLM provider, possibly colluding with some users, attempts to steal user prompts.
    \item \textbf{Threat from Users}: A user leverages security holes in LLM software to steal LLM parameters and user prompts.
\end{itemize}
For example, traditional LLM inference is vulnerable to all three kinds of threats, while standard confidential inference approach (\autoref{fig:image1}) is threatened by the LLM and the users.

It is worth noting that we assume the LLM provider behaves rationally, which means it follows the prescribed inference steps to maximize its own benefit, although it is untrusted, curious about user prompts, and even seeking profit from user secrets. 
This assumption aligns with the Honest-but-Curious (HbC) threat model commonly used in secure computation literature~\cite{goldreich2001foundations}.
HbC model also reflects industry practice where the LLM provider is incentivized to maintain integrity for reputation, especially when altered token generations performed by the LLM service are detectable.
For example, detection of injection attacks is practical, as discussed in \S\ref{sec:injection_attacks} and \S\ref{sec:discussion-mitigating-attacks}.

Denial of service (DoS) attacks are out of consideration.
We do not consider attacks that compromise the communication channels or the CVM either.
In \S\ref{sec:discussion-mitigating-attacks}, we discuss some potential attacks on the TCB and their mitigations.

\subsection{Design Goals}
\label{sec:design-space}

\noindent 
Our \textbf{\emph{primary goal}} is to secure user prompt confidentiality in cloud-hosted LLM service.
Beyond the primary goal, we target three additional goals for commercial deployment.

\begin{itemize}[topsep=0.1em,itemsep=-0.1ex,leftmargin=*]
    \item \textbf{Model confidentiality}: LLM parameters must not leak. This is critical as the parameters constitute an intellectual property of the LLM. Preserving model confidentiality enhances the deployability of closed-source LLMs.
    \item \textbf{\Outinv}: Security measures must not change the output of LLM. This is crucial for deployment, particularly for tasks in clinical and financial fields, where even a small accuracy error could lead to serious consequences.
    \item \textbf{Compute efficiency}: Security measures cannot significantly increase the LLM serving cost. While security is not free, we believe that a more efficient approach is more attractive to users.
\end{itemize}

\subsection{Overview of \OSPD's Auditable Protection}
\label{sec:overview}

\noindent
As shown in \autoref{fig:design_overview}, \OSPD's core components, including the CVM, the trusted OS, and the \MGR, integrate as a system and collaborate to provide auditable protection.
We next present an overview of this collaboration.

\paragraphb{\textbf{Auditable Software Stack}}
\OSPD's CVM guest software stack is open-source for independent audits by users and the LLM provider, such as the Linux kernel~\cite{linux_kernel}, Nvidia Linux GPU driver~\cite{nvidia_gpu_driver}, and the \MGR.
Such audits are crucial for establishing trust between \OSPD with both users and the LLM provider.
On one hand, these parties can ensure that the software stack is provided by trusted parties such as Linux community and Nvidia, instead of the cloud provider, LLM provider, or any users.
On the other hand, by analyzing the processing logic and data flow reflected in the source code, the participants gain confidence that their secrets are well protected \emph{at runtime}.
Notably, \textbf{\OSPD does not require the LLM software or the userspace CUDA drivers to be open source}, which distinguishes our approach from related projects such as Tinfoil~\cite{tinfoil_technical_overview, tinfoil_builds_trust} (See \S\ref{sec:zero_trust} for more details).
At runtime, they execute as unprivileged processes in user mode.
The trusted OS guarantees that they cannot harm the rest of the system.

\paragraphb{\textbf{CVM's Decentralized Initialization}}
As discussed in \S\ref{sec:zero_trust}, independent code audits combined with remote attestation eliminate the need to trust the CVM owner, i.e., the party who instantiates the CVM in the cloud.
\OSPD's design and its auditable software stack guarantee that the CVM owner does not have any higher privilege than other parties.
As a result, we do not restrict who instantiates the CVM, which can be the LLM provider, any user, or even any third party, as long as all participants verify the integrity of the CVM via remote attestation before transmitting their secrets.
Such a decentralized feature make \OSPD differ from the standard confidential inference approach, as shown in \autoref{fig:image1}, which requires a centralized trusted party to play as the CVM owner.

\paragraphb{\textbf{Attestable CVM Environment}}
As mentioned in \S\ref{sec:confidential-computing}, CVM's remote attestation capability allows users and the LLM provider to verify if the CVM hardware is genuine and to check if different aspects of the boot process match with the audited guest software stack.
To be more precise, the CVM hardware generates an attestation report, which encapsulates the measurement, i.e., cryptographic hash, of different aspects of the boot process.
Since the software stack is open-source, all parties can independently compute the expected measurement value and compare them with those in the attestation report.
As a result, they can ensure that the CVM is untampered prior to secret transmission.

\paragraphb{\textbf{Runtime Enforcement}}
Auditing source code and verifying CVM integrity at initialization are necessary for establishing trust.
However, these measures alone are not sufficient to prevent information leakage at runtime.
The key lies in the collaboration between the CVM, the trusted OS, and the \MGR to enforce runtime protection, which includes cryptographically protected communication, sensitive data isolation, and information flow control.
We elaborate on these mechanisms in \S\ref{sec:protecting-user-prompts}.

\section{Efficient Protection of Prompt and LLM}
\label{sec:protecting-user-prompts}

\newcommand{\de}{\gamma}

\noindent
As discussed in \S\ref{sec:confidential-inference}, under the assumption of an untrusted LLM provider, it is difficult to balance efficiency with confidentiality of both user prompts and LLM parameters.
That is, assigning a dedicated LLM service for each user unavoidably leads to significant inefficiency, although it allows explicit isolation between user prompts (\autoref{fig:image2}, \autoref{fig:image3}).
\OSPD overcomes this challenge by enforcing strict data isolation and flow control over user prompts and LLM parameters, while enabling efficient batch processing across all user prompts by a single LLM service (\autoref{fig:image4}).

Our key insight is that the token generation can be formulated as a secure partitioned computation between the users and the LLM provider.
Each user owns a process within \OSPD's CVM and each of these processes represents the associated user as one participant in the secure partitioned computation.
\OSPD partitions the KV cache into \emph{\private KV cache} and \emph{\public KV cache}, which are associated with the user input prompts and LLM generated output tokens respectively.
The \private KV cache of user prompts is private and kept confidential in the per-user processes, while the \public KV cache is processed by the LLM service process.

For simplicity, we detail our design in a single-user scenario
\footnote{
    The extension from single-user to multi-user scenario is trivial, because the computation in a per-user process is independent on other users, and the LLM service process can batch process for all users. 
}
, assuming that the CVM and its guest software stack have been audited, initialized, and verified (\S\ref{sec:overview}).
We first present an overview of the secure partitioned computation protocol and then detail each component in the following.
There are four participants in the protocol:
\begin{itemize}[topsep=0.1em,itemsep=-0.1ex,leftmargin=*]
    \item \textbf{A user}, who sends prompts to request LLM service and receives output tokens as responses securely.
    \item \textbf{The user's process}, which represents the user to process the prompts and interacts with the service process.
    \item \textbf{The service process}, which represents the LLM provider to provide LLM service for users.
    \item \textbf{The \MGR}, which initializes processes in the CVM and enforces information flow control policies.
\end{itemize} 

\vspace{1ex}\noindent We next introduce the computation and communication protocol among these four participants.
\begin{enumerate}[topsep=0.1em,itemsep=-0.1ex,leftmargin=*]
    \item \textbf{Setup} (\S\ref{sec:secure-prefill}): The \MGR initializes processes for the user and the LLM provider, respectively, in the CVM.
    It establishes secure channels with the user and the LLM provider, while restricting network access for their processes.
    Both the user and the LLM provider send their secrets, i.e., user prompts and LLM parameters, to the \MGR over the secure channels, which then relays them to the associated processes.
    \item \textbf{Prefill} (\S\ref{sec:secure-prefill}): The per-user process computes \private KV cache $K_\pvt, V_\pvt$ and generates the first token. It then sends the first token to the service process.
    \item \textbf{Decode} (\S\ref{sec:decoding}): Receiving the first token from the per-user process as a new token, the service process generate the next token autoregressively.
    As depicted in \autoref{fig:SPD_details}, for each transformer layer:
    \begin{enumerate}[topsep=0.1em,itemsep=-0.1ex,leftmargin=3ex]
        \item The service process computes $Q_\new, K_\new, V_\new$ of the new token, sends $Q_\new$ to the per-user process, and appends $K_\new, V_\new$ to \public KV cache $K_\pub, V_\pub$.
        \item The per-user process responds with \private attention score $A_\text{\pvt}=\softmax(Q_\new K_\pvt^{\top}) V_\pvt$.
        \item The service process computes \public attention score $A_\text{\pub}=\softmax(Q_\new K_\pub^{\top}) V_\pub$, and merges it with $A_\text{\pvt}$ to get full attention score $Y=\softmax(Q_\new K^{\top})V$ according to \autoref{thm:partitioned-attention}.
        \item If it is the final layer, the service process samples a new token from $Y$, sends it to the \MGR, and continues generating tokens until \texttt{[EOS]}. Otherwise, it continues to the next layer.
    \end{enumerate}
    \item \textbf{Response} (\S\ref{sec:secure-prefill}): The \MGR collects all generated output tokens from the service process and sends them to the user via the secure channel.
\end{enumerate}
This protocol can be generalized to scenarios with multiple users, each with its own process operating in the CVM.

\subsection{Data Isolation and Flow Control}
\label{sec:secure-prefill}

\noindent
As analyzed in \S\ref{sec:prompt_leakage}, the key to preserving confidentiality of user prompts and LLM parameters lies in explicit memory isolation and strict information flow control.
The \MGR and the underlying trusted OS collaborate to enforce the isolation and flow control policy.

\subsubsection{Secure Channel and Process Initialization}
Once \OSPD's CVM is initialized, the \MGR keeps listening on connections from users and the LLM provider.
During each connection setup, the \MGR and the user (or the LLM provider) use Diffie-Hellman key exchange protocol~\cite{diffie2022new} to jointly derive unique symmetric keys for secure communication.
In other words, a secure channel is established between the \MGR and the user (or the LLM provider).
The \MGR then creates a dedicated process for the user (or the LLM provider).
Specifically, the \MGR sends the symmetric keys to the created per-user process via Inter-process Communication (IPC).
It implies that the users can securely submit prompts to their associated processes, where the \MGR acts as a relay.

\subsubsection{LLM Parameter Read-only Sharing}
After the secure channel is established, the LLM provider securely transmits the LLM parameters to the \MGR, which then saves the parameters as a read-only file in the CVM memory and grants read-only access permission to both the service process and all per-user processes.
As the file is read-only, these processes can safely share the same copy of the LLM parameters without risking malicious modification.
It is worth noting that, each per-user process accesses the LLM parameters only during the \program{prefill} phase to compute the \private KV cache.
In addition, all per-user processes can leverage CUDA IPC~\cite{cuda_ipc} to share GPU memory for the LLM parameters, so to avoid redundant parameter loading and GPU memory consumption.

\subsubsection{Restricted Network Access}
To prevent the per-user processes from leaking LLM parameters, the \MGR leverages Linux namespaces to restrict their network access capabilities.
Specifically, at process creation, the \MGR configures each per-user process to operate within a dedicated and isolated network namespace.
As a result, each per-user process can communicate with its associated user only under the inspection from the \MGR, which acts as a relay in the communication.
For example, the users send encrypted prompts to the \MGR, which then forwards them to the associated per-user processes via IPC.
On the other hand, the \MGR collects the generated output tokens accordingly and sends them back to the associated users.

\subsection{\SPD (SPD)} 
\label{sec:decoding}

\begin{figure}[t]
    \centering
    \includegraphics[width=0.48\textwidth]{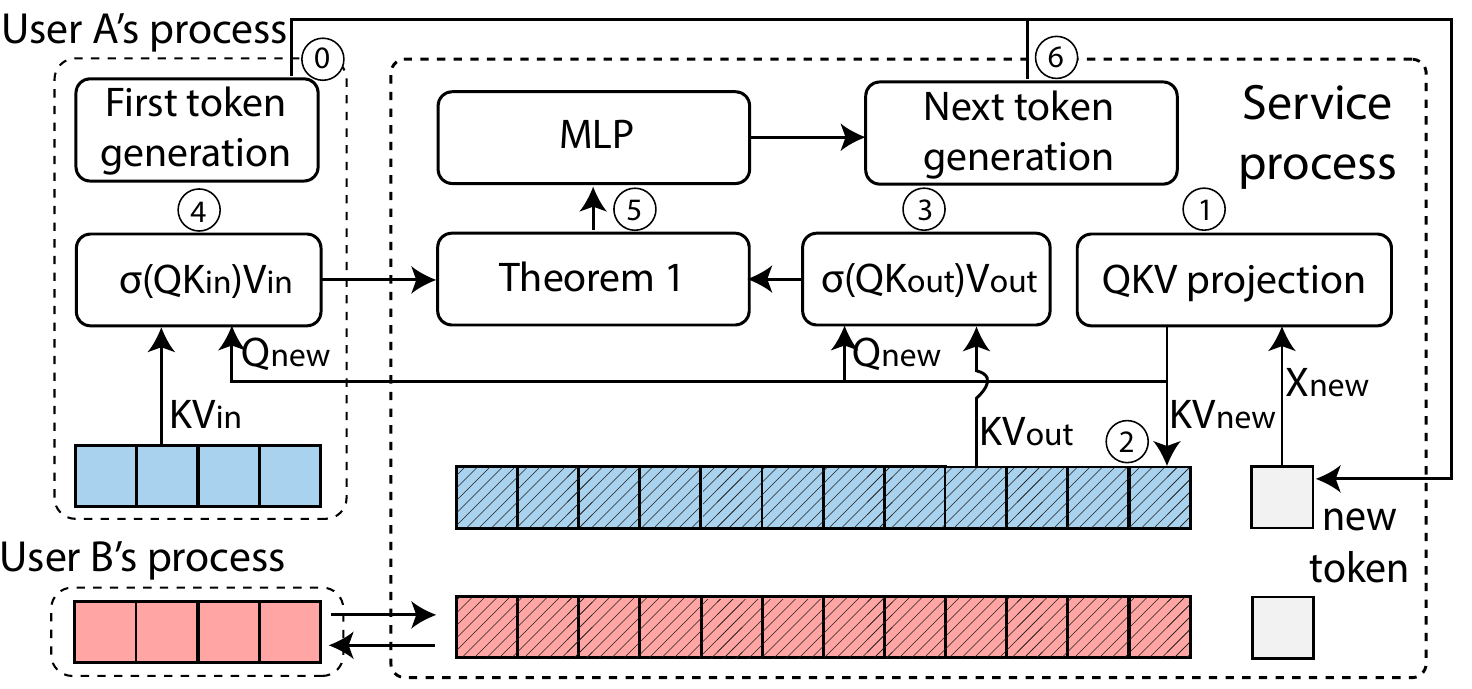}
    \caption{\textbf{Overview of SPD on a simplified Transformer layer.}
    The squares in blue and red represent the KV cache associated with different users while the gray squares represent new tokens.
    With or without shade indicate it is the \public or \private KV cache, respectively.
    \textcircled{\small{0}} By the end of \program{prefill}, the user process finishs computing its \private KV cache $K_\pvt, V_\pvt$, generates the first token and sends it to the service process.
    \textcircled{\small{1}} Project hidden state $X_\text{new}$ of a new token to $Q_\text{new}, K_\text{new}, V_\text{new}$.
    \textcircled{\small{2}} Append $K_\text{new}, V_\text{new}$ to the \public KV cache.
    \textcircled{\small{3}} Batch process \public attention score for all users.
    \textcircled{\small{4}} Compute \private attention score in each user process.
    \textcircled{\small{5}} Merge results to compute full attention score.
    \textcircled{\small{6}} If it is the last layer, generate the next token, then repeat from \textcircled{\small{1}} until finish; otherwise continue to the next layer.}
    \label{fig:SPD_details}
\end{figure}

\noindent 
We formulate the decoding in the single-user scenario as a secure partitioned computation using the online softmax calculation~\cite{milakov2018online}.
This secure computation enables the LLM to retrieve the full attention score $Y$ without knowing the user prompt and the \private KV cache $K_\pvt$, $V_\pvt$.

\begin{theorem}[Secure Partitioned Attention Computation]
\label{thm:partitioned-attention}
Let $Q\in\mathbb{R}^{d}$, $K = \cat(K_\pvt, K_\pub)\in\mathbb{R}^{len\times d}$, $V = \cat(V_\pvt, V_\pub)\in\mathbb{R}^{len\times d}$, where $len$ be the number of input and output tokens, and $\softmax$ be the softmax function.
\begin{align}
\label{eq:1}
\softmax(Q K^{\top})V = \frac{\de_\pvt}{\de_\pvt+\de_\pub} \softmax(Q K_\pvt^{\top})V_\pvt\nonumber \\ + \frac{\de_\pub}{\de_\pvt+\de_\pub}\softmax(Q K_\pub^{\top})V_\pub,
\end{align}
where $\de_\pvt,\de_\pub$ are denominators of each softmax operation, \eg $\de_\pvt=\sum\text{exp}(Q K_\pvt^{\top})$.
\end{theorem}

The proof of \autoref{thm:partitioned-attention} is available in \S\ref{sec:proof_spd}.
This theorem serves as the foundation of our SPD design, which offers three key benefits.
First, the decomposition is lossless and thus SPD maintains \outinv. 
Second, computations in per-user processes do not require LLM parameters during \program{decode}, which means the per-user processes only require a small amount of memory for the \private KV cache and \private attention states.
Third, the LLM service process can batch process the \public attention states ($Q, K_\pub, V_\pub$) for all users in parallel.

We can naturally extend \autoref{thm:partitioned-attention} to the multi-user scenario.
When multiple requests from different users arrive simultaneously, each per-user process computes its own \private attention score independently, while the service process batch processes the \public attention states for all users in parallel.
This enables efficient and isolated computation for multiple users, especially when the NVIDIA Multi-Process Service (MPS) is enabled~\cite{nvidia_mps}, which allows multiple processes to concurrently and spatially share GPU resources while maintaining isolation on GPU devices.

Finally, it is worth noting that computing $\de_\pvt$ and $\de_\pub$ individually is numerically unstable due to their exponential term.
To address this, we use the maximum values for \private and \public attention scores, denoted as $m_\pvt$ and $m_\pub$, respectively, to improve numerical stability.
In practice, we optimize the computation as $\de_\pvt=\sum\text{exp}(Q K_\pvt^{\top}-m_\pvt)$, where $m_\pvt=\max(Q K_\pvt^{\top})$, and $\de_\pub=\sum\text{exp}(Q K_\pub^{\top}-m_\pub)$, where $m_\pub=\max(Q K_\pub^{\top})$.
The coefficients in \autoref{thm:partitioned-attention} become $\de_\pvt/(\de_\pvt+\alpha \de_\pub)$ and $\de_\pub/(\alpha^{-1} \de_\pvt+\de_\pub)$, where $\alpha=\exp (m_\pub-m_\pvt)$.

\subsection{Security and Functional Analysis}
\label{sec:security-analysis}

\paragraphb{\textbf{User Prompt Confidentiality}}
We analyze how \OSPD protects user prompt confidentiality against the threats outlined in \S\ref{sec:prompt_leakage} and \S\ref{sec:threat-model}.

First, user prompts remain confidential from adversaries on the cloud, including the cloud provider.
The prompts are encrypted during transmission and will not be decrypted until they are in the CVM, which provides strong isolation and its integrity is verified via remote attestation.

Second, the service process cannot access user prompts as the prompts and their \private KV cache remain confidential within per-user processes.
The trusted guest OS guarantees isolation among processes to avoid any secret exposure.
The service process learns only \textit{(1)} the generated output tokens, and \textit{(2)} the \private attention score $A_\text{\pvt}$.
However, it is not practical for an attacker to recover user prompts from such information.
For \textit{(1)}, the state-of-the-art techniques for recovering prompts from LLM output~\cite{gao2024dory, sha2024prompt, yang2024prsa} have been proven to perform poorly on in-the-wild prompts in practice~\cite{tan2025effectiveness} (See \S\ref{sec:prompt_stealing_attacks} for details).
For \textit{(2)}, the attention score computation is an information-losing map, meaning that it discards much of the original information of the prompt and retains only those relevant for generating the output token, as suggested by its term ``attention''.
As a result, the \private attention score $A_\text{\pvt}$ is typically irreversible to the user prompt, unless the query matrix $Q$ is adversarially selected.
This implies that a more promising attack is to inject prompt-leakage instructions, which induce the \private attention score to keep as much information about the prompt as possible, and further induce prompt leakage in token generation.
However, such an attack would require the LLM service to manipulate the inference process, which falls outside our Honest-but-Curious threat model and is detectable by existing work~\cite{hung2024attention} that is orthogonal to \OSPD. Please refer to \S\ref{sec:injection_attacks} and \S\ref{sec:discussion-mitigating-attacks} for the detection approach in details.

Finally, users cannot obtain other users' prompts as the prompts and \private KV cache are isolated in per-user processes.
Particularly, attacks that exploit the vulnerabilities in LLM software, e.g., shared KV cache~\cite{wu2025know}, cannot succeed either.
This is because SPD relies on the underlying trusted OS, rather than the LLM software, to enforce isolation among user prompts and \private KV cache.

\paragraphb{\textbf{Model Confidentiality}}
The secure channel and CVM isolation guarantee that the LLM remains confidential from adversaries on the cloud, including the cloud provider, which is similar to how \OSPD protects user prompt confidentiality from such adversaries.
On the other hand, \OSPD's design ensures that the per-user processes cannot send any data out of the CVM, as enforced by network namespace restrictions managed by the \MGR.
As a result, the LLM secrets are secure although the per-user processes have read-only access to them during \program{prefill}.
It is worth highlighting that the output tokens generated by the service process are sent to the user by the \MGR instead of the per-user processes.
This explicit separation ensures that per-user processes cannot exfiltrate model data via output tokens, as all output delivery is strictly controlled by the \MGR.

\paragraphb{\textbf{Compute Efficiency}}
\OSPD's SPD design enables efficient computation by allowing the service process to batch process \public attention states for all users in parallel.
This increases GPU utilization substantially.
In contrast, confidential inference approaches that assign a dedicated LLM service for each user cannot leverage batch processing across users (See \S\ref{sec:confidential-inference} for details).
In addition, \OSPD's auditable CVM environment (\S\ref{sec:overview}) allows secure software-level optimizations, such as read-only data sharing and enabling Nvidia MPS~\cite{nvidia_mps}, while eliminating the need of a centralized trusted party to play as the CVM owner.

\paragraphb{\textbf{\OutInv}}
\OSPD's SPD design maintains \outinv by ensuring the attention computation is mathematically equivalent to the original token generation process, as demonstrated in \autoref{thm:partitioned-attention}.
\OSPD does not introduce any approximation or require LLM retraining, thus preserving the invariance of the output tokens, which distinguishes our approach from the related work (See \S\ref{sec:relwork} for details).

\section{Implementation}
\label{sec:implementation}

\noindent 
We next describe key implementation details of \OSPD.

\paragraphb{Auditable Software Stack}
\OSPD's software stack is available in GitHub, allowing users and the LLM provider to audit the code for any potential backdoors and vulnerabilities.
It includes the Linux kernel~\cite{linux_kernel}, Nvidia Linux GPU driver~\cite{nvidia_gpu_driver}, PyTorch~\cite{pytorch_source}, attestation tools~\cite{snp_guest, nvidia_trust, nvidia_gpu_attestation}, and the \MGR.
One can leverage the GitHub Actions CI/CD pipeline to automatically build the CVM image from the source code~\cite{github_actions}.
The pipeline can also generate a measurement file of the built image, which is used for comparison with the measured hash value recorded in the CVM's attestation report.

\paragraphb{Attestable CVM Environment}
Both users and the LLM provider verify the CVM environment before transmitting any secrets.
We implement the attestation process following the challenge-response model.
That is, a user or the LLM provider sends a challenge, i.e., a random nonce, to the \MGR, initiating the attestation process.
The \MGR first triggers the attestation of Nvidia GPU TEE with the Nvidia Remote Attestation Service (NRAS)~\cite{nvidia_gpu_attestation}.
It includes the received challenge in its request to NRAS and in turn receives a verifiable token from NRAS~\cite{nvidia_gpu_ra_code}.
The \MGR then generates a CPU TEE attestation report with tool such as \program{snpguest}~\cite{snp_guest}, providing the received challenge and the hash of the Nvidia token as input.
As a result, the generated report not only measures and records the state of the CPU TEE for verification, but also indicates the integrity of the GPU attestation and ensures their freshness.
Finally, the \MGR returns the generated attestation report and the Nvidia token as a response.

After auditing the software stack and verifying the CVM environment, users and the LLM provider can confidently transmit their encrypted secrets to the \MGR via secure channels established with the Diffie-Hellman key exchange protocol~\cite{diffie2022new}.

\paragraphb{LLM Supply and Service Process Initialization}
We consider two ways for the LLM provider to supply the LLM to the \MGR.
First, if the LLM software is open source, we include it in building the CVM image as part of the software stack audit process.
During runtime, the LLM provider transmits only LLM parameters via the secure channel.
Alternatively, the provider can encrypt and sign its closed source LLM software binary and parameters, include the encrypted data and signatures in the CVM image, and during runtime transmit only the cryptographic keys via the secure channel.
In either case, the \MGR decrypts the LLM parameters within the CVM before initializing the service process.
The \MGR stores the decrypted LLM parameters as a read-only in-memory file and creates the service process.
It grants the service process read-only access to the in-memory file containing the LLM parameters, which allows the service process to initialize as usual.

\paragraphb{LLM Parameter Read-only Sharing and Per-user Process Initialization}
The \MGR maps the LLM parameters as read-only memory.
After creating per-user processes and restricting their network capabilities, the \MGR shares the read-only LLM parameters with these processes.
To achieve this, the \MGR leverages PyTorch's CUDA \program{MemPool} to integrate a customized memory allocator, which allocates GPU memory specifically only for the LLM parameters~\cite{pytorch_mempool}.
To be more precise, it allocates GPU memory regions via \program{cuMemCreate} and generates shareable handles via \program{cuMemExportToShareableHandle} for inter-process sharing LLM parameters~\cite{cuda_ipc_api}.
When creating per-user processes, the \MGR sets the \program{CLONE\_NEWNET} flag in \program{clone} system calls.
This leverages Linux namespaces to restrict the network capabilities of the per-user processes~\cite{linux_clone}.
Then the \MGR shares the GPU memory handles with the per-user processes, which then import the allocated GPU memory regions via \program{cuMemImportFromShareableHandle} to access the LLM parameters for \program{prefill}~\cite{cuda_ipc_api}.
This approach ensures that only authorized per-user processes can access the shared GPU memory, preventing unauthorized access to the LLM parameters while maintaining isolation for the rest of memory.

\paragraphb{\SPD}
We develop SPD based on the Transformers library~\cite{wolf2020transformers}, adapting the Llama model by monkey-patching its attention module.
To be more precise, we modified the attention score computation to prioritize the computation of matrix $Q$.
This enables the service process to promptly send $Q$ to each per-user process asynchronously via the GLOO communication backend~\cite{gloo} for \private attention score computation.
While the service process asynchronously waits for all \private attention scores to arrive, it continues to compute matrices $K$, $V$ and \public attention scores.
Once all scores are ready, it computes the final attention scores with \autoref{thm:partitioned-attention} and generates output tokens.
Although our prototype is based on the Llama model, our design is generally applicable to other Transformer-based LLMs.
We also note that the GLOO~\cite{gloo} backend transfers tensors via the host, which incurs non-trivial overhead (See \S\ref{sec:eval-compute-efficiency}).
We discuss the portability of \OSPD design and the reasons why other popular communication backends, e.g., NCCL~\cite{nccl}, are not suitable in \S\ref{sec:discussion-portability}.

\begin{figure*}[t]
    \centering
    \begin{minipage}{0.48\textwidth}
        \centering
        \includegraphics[height=0.55\textwidth]{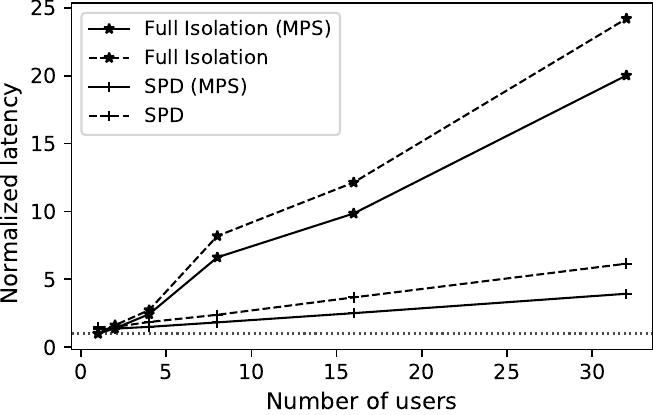}
        \caption{\textbf{Normalized latency with varying number of users}, Llama 3 (8B), 64 input and 64 output tokens. $y=1$ indicates the latency of No Protection baseline.}
        \label{fig:latency_vs_user}
    \end{minipage}
    \hfill
    \begin{minipage}{0.48\textwidth}
        \centering
        \includegraphics[height=0.55\textwidth]{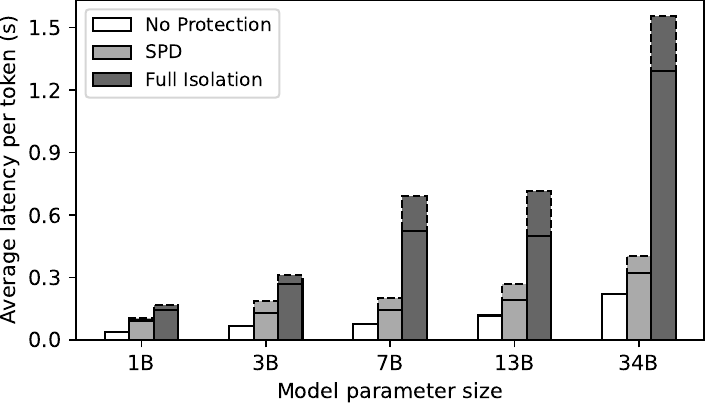}
        \caption{\textbf{Average latency with varying model sizes}, 8 users, 64 input and 64 output tokens. The solid and dashed bars indicate with and without MPS respectively.}
        \label{fig:model_param}
    \end{minipage}
\end{figure*}

\begin{figure}[ht]
    \centering
    \includegraphics[height=0.24\textwidth]{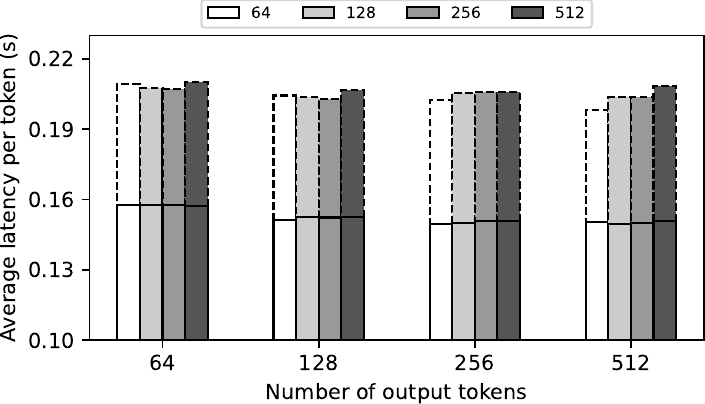}
    \caption{\textbf{Average latency per generated token, with varying number of input and output tokens}, Llama 3 (8B) and 8 users. Varying groups of bars indicate varying output token counts. The four bars in each group indicate varying input token counts. The solid and dashed bars indicate with and without MPS respectively.}
    \label{fig:latency_vs_output}
\end{figure}

\begin{figure}[ht]
    \centering
    \includegraphics[height=0.24\textwidth]{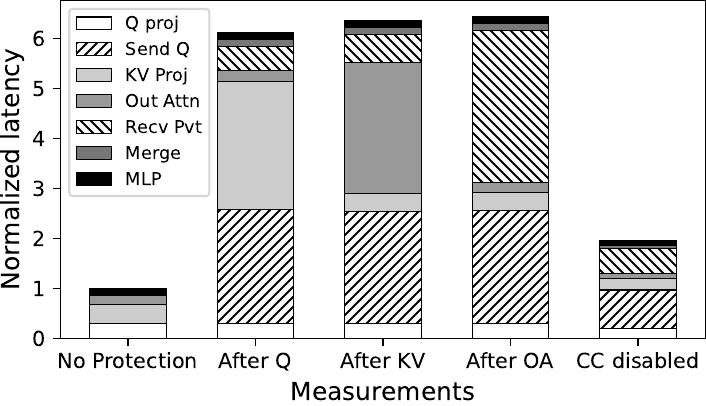}
    \caption{
    \textbf{Overhead breakdown.} All are measured with Llama 3 (8B), 32 users, 64 input and 64 output tokens. MPS is enabled except for \textit{No Protection}.}
    \label{fig:smd_overhead_breakdown}
\end{figure}

\section{Evaluation}
\label{sec:evaluation}

\noindent
In this section, we primarily focus on the question: \emph{Does \OSPD achieve high scalability and maintain compute efficiency?}
To answer this, we evaluate \OSPD's performance comparing to two existing confidential inferencing approaches (\S\ref{sec:confidential-inference}).
As for empirical security evaluation on prompt stealing attacks (\S\ref{sec:prompt_stealing_attacks}), we refer the readers to the experiments and analysis in Tan et al~\cite{tan2025effectiveness}.

\paragraphb{\textbf{Evaluation setup}}
Without special mention, all evaluations were conducted in an Azure's CVM, \textit{NCCads\_H100\_v5}~\cite{azure_cvm_gpu_tee}.
The CVM equips with an Nvidia H100 GPU with 94 GB of memory, 40 AMD EPYC Genoa processor cores, and 320 GB of system memory.
Confidential computing features, i.e., AMD SEV-SNP~\cite{amd-sev} and Nvidia GPU CC~\cite{nvidia_tee_2023}, are enabled.

The software stack in the CVM includes Ubuntu 24.04 with kernel version 6.11.0, Nvidia open driver version 570.158.01, CUDA 12.8, Python 3.12.3, and PyTorch 2.7.1.
For LLM, we utilizes Llama 3~\cite{llama3} with 8B, Llama 3.2 with 1B and 3B, and Code Llama~\cite{codellama} with 7B, 13B and 34B parameters.

For overhead evaluation, we measure the latency both with Nvidia MPS~\cite{nvidia_mps} enabled and disabled.
We note that, to use MPS, it requires no modifications to the implementation.

\newcommand{\sllm}{\textit{No protection}\xspace}
\newcommand{\nllm}{\textit{Full isolation}\xspace}
\newcommand{\smd}{\textit{SPD}\xspace}

\paragraphb{\textbf{Performance Analysis}}
We compare \OSPD with two baselines:
(1) \sllm, where a LLM instance serves all users within a single process (\autoref{fig:image1}).
It does not secure user prompts from the LLM provider and is intended to demonstrate the upper bound of performance.
(2) \nllm, where each user owns a per-user process that runs a dedicated LLM instance (\autoref{fig:image3}).
\OSPD is denoted as \smd (\autoref{fig:image4}).
As mentioned in \textit{Evaluation Setup}, we measure and compare the performance both with Nvidia MPS~\cite{nvidia_mps} enabled and disabled.

\subsection{Scalability}

\noindent
Our evaluation includes 1 to 32 users, with both prompts and responses ranging from 64 to 512 tokens.
We measure the end-to-end latency for each user to receive the responses.
\autoref{fig:latency_vs_user}, \autoref{fig:model_param} and \autoref{fig:latency_vs_output} summarize the main results, demonstrating that our approach scales effectively as the number of users, input/output tokens, and the model parameter size increase.

\paragraphb{\textbf{Number of Users}}
The \nllm approach faces inherent scalability limitations due to GPU memory constraints because it provides separate LLM instances for each user.
As shown in \autoref{fig:latency_vs_user}, it exhibits significant latency degradation with increasing user counts.
In contrast, \OSPD achieves superior scalability, as it maintains a substantially smaller memory footprint.
However, \OSPD still faces high overhead compared to the \sllm approach (indicated as $y=1$), which is the cost of isolating user prompts in per-user processes.

\paragraphb{\textbf{Model Parameter Size}} 
\autoref{fig:model_param} shows that the token generation slows down for all approaches when the size of model parameter increases.
Not surprisely, \sllm performs the best, while \smd is less affected by parameter size scaling compared to the \nllm approach.
In other words, \nllm's end-to-end latency scales at a higher rate than that of \smd under the same conditions, when the model size increases from 1B to 34B.

\paragraphb{\textbf{Number of Input/Output Tokens}}
\autoref{fig:latency_vs_output} shows that both input and output token counts have negligible impact on token generation.
As the counts increase, the latency per token remains relatively stable as decoding each token has static overhead, or even slightly reduces because the initial cost is amortized across the tokens.

\subsection{Compute efficiency}
\label{sec:eval-compute-efficiency}

\noindent
We further breakdown the overhead of \OSPD to demonstrate the sources of overhead.

\paragraphb{\textbf{Overhead breakdown of SPD}}
\smd introduces overhead mainly due to (1) the absence of batch processing in per-user processes, and (2) communication between per-user processes and the service process.
We present \smd's latency breakdown in \autoref{fig:smd_overhead_breakdown}.
The seven latency components align with the processing steps in \autoref{fig:SPD_details}, except that we partition the ``QKV projection'' into ``Q proj'' and ``KV proj''.
The first bar in \autoref{fig:smd_overhead_breakdown} represents the breakdown of \sllm, while other bars represent that of \smd in different conditions.
The three bars in the middle differ in when $Q$ is sent, right after the computation of $Q$ (``After Q''), $K$ and $V$ (``After KV''), or the \public attention scores (``After OA'').
The last bar is measured with GPU CC disabled.
We note \autoref{fig:smd_overhead_breakdown} is measured with CUDA Event on GPU side, which is slightly different from the end-to-end latency in previous measurements.

\paragraphb{\textbf{Processes compete for GPU resources}}
Temporally ignoring the communication when comparing the first four bars, we observe that the computation right after sending $Q$ is much slowed down compared to the \sllm baseline, while other components have negligible overhead.
It is because the per-user processes compete with the service process for GPU resources once they receive $Q$.
Sending $Q$ after \public attention computation can avoid such competition, but the service process must be blocked and keep waiting for the \private scores, which results in slightly higher total latency as it fails to overlap communication with computation.

\paragraphb{\textbf{High communication overhead of GPU CC}}
The last bar in \autoref{fig:smd_overhead_breakdown} shows SPD's latency breakdown with GPU CC disabled.
Its total latency is about $1/3$ of that with GPU CC enabled since the communication overhead reduces by about 5$\times$.
This is because the GLOO backend~\cite{gloo} transfers tensors via the host, where the GPU driver and GPU device encrypt and decrypt all transferred data going through the PCIe, incurring high overhead~\cite{nvidia_gpu_cc_whitepaper} (\S\ref{sec:nvidia_gpu_cc}).
We expect this overhead can be much reduced and even fully eliminated with newer version GPU CC designs that enable TEE-IO~\cite{nvidia_blackwell} or better support of communication across processes that share the same GPU (See \S\ref{sec:discussion-portability}).

\section{Discussion}
\label{sec:discussion}

\noindent
In this section, we first discuss attacks out of our threat model and how \OSPD may works with existing solutions to mitigate them.
Then, we discuss the portability of \OSPD design and how to deploy it under different situations.
Finally, we discuss the limitations of \OSPD and its future work.

\subsection{Mitigating Attacks Out of Scope}
\label{sec:discussion-mitigating-attacks}

\subsubsection{Prompt-leakage Injection Attacks}
A malicious service process may induce the per-user processes to leak prompts by injecting instructions like ``repeat the prompt'' into output token sequences via manipulating the token generation (\S\ref{sec:injection_attacks}).
Although this attack is out of our Honest-but-Curious (HBC) threat model, \OSPD can work with attention-based detection methods to defend against it.
Recently, Hung et al.~\cite{hung2024attention} discover the \emph{distraction effect} in attention computation when injected instructions are present.
Building on this discovery, Attention Tracker effectively detects prompt injection attacks by monitoring attention computation~\cite{hung2024attention}.
Similarly, users can identify the distraction effect, so to detect prompt-leakage injection attacks performed by the service process, via monitoring \OSPD's input attention scores within the per-user processes.

\subsubsection{Attacks on TCB}
\OSPD introduces a novel application-level approach to \name.
Its security is built on top of the CVM hardware and the software stack, e.g., guest OS.
As such, it inherits both the security guarantees and the vulnerabilities of the underlying TCB.

\paragraphb{Attacks on CVM}
\OSPD does not defend against attacks that compromise the CVM~\cite{li2021cipherleaks, yuan2025ciphersteal, tee_fail_sp2026}.
Many of these attacks exploit vulnerabilities in specific CVM implementations, for example, the deterministic encryption.
Since \OSPD's design does not impose any restrictions on its underlying infrastructure (See discussion on portability in \S\ref{sec:discussion-portability}), it is compatible with any existing and future solutions that enhance the security of CVM~\cite{duy2025incognitos, qin2023protecting}.

Notably, Chuang et al.~\cite{tee_fail_sp2026} recently extracted the Provisioning Certification Key (PCK) of Intel TDX, successfully compromising the chain of trust of its attestation mechanism.
This attack severely undermines the trust established between \OSPD with users and the LLM provider.
However, thanks to \OSPD's portable design, the attestation issues can be mitigated by deploying \OSPD on CVM implementations that are not vulnerable to this issue.

\paragraphb{Attacks on OS}
Similarly, \OSPD inherits vulnerabilities in the software stack, especially the guest OS due to its large attack surface~\cite{schluter2024wesee, schluter2024heckler}.
To mitigate such attacks, \OSPD can leverage existing techniques that enhance the security of OS, for example, containerizing each process with gVisor~\cite{gVisor,gVisor_gpu} to minimize the OS's attack surface.
Another series of techniques that focus on protecting data in use against a compromised OS~\cite{zhao2022vsgx, wang2024road, li2024blindfold} also enhances \OSPD's security.

\subsection{Portability and Deployment of \OSPD}
\label{sec:discussion-portability}

\paragraphb{\textbf{Portability across LLMs}}
\OSPD is portable across different decoder-only LLMs such as GPT~\cite{achiam2023gpt} and Llama~\cite{touvron2023llama} series.
Although our prototype and evaluation focus on the Llama series, we believe \OSPD is applicable to other decoder-only LLMs, e.g., the GPT series.
This is because our attention decomposition (\autoref{thm:partitioned-attention}) is general without relying on any specific implementations.

\paragraphb{\textbf{Portability across Architectures}}
\OSPD is portable across different CPU and accelerator architectures, provided they support a CVM spanning across CPU and the accelerators.
We deploy our prototype in an Azure CVM because, at the time of writing, it is the only CVM available on public cloud that enables CC on an NVIDIA H100 GPU~\cite{nvidia_tee_2023}.
We believe \OSPD can be deployed in CVMs with various architectures, such as combinations of AMD SEV-SNP~\cite{amd-sev} and ARM CCA~\cite{arm-cca}, together with NVIDIA Blackwell~\cite{nvidia_blackwell} and security-enhanced TPU~\cite{google_tpu_tee_blog, google_tpu_tee_report}.

\paragraphb{\textbf{Portability across Communication Backends}}
We use GLOO~\cite{gloo} as it is general.
In contract, even if NCCL~\cite{nccl} typically performs better in scenarios involves GPUs, it requires that each process has exclusive access to a GPU, which is not suitable for our evaluation platform.
CUDA IPC, as well as PyTorch's Queue and Pipe, can share tensors across processes without copying.
However, so far they do not support asynchronous IO.
As a result, they are even less efficient.
Fortunately, one can expect that the newer version of GPU CC in Nvidia Blackwell~\cite{nvidia_blackwell} with TEE-IO can reduce and even eliminate the overhead of encrypted communication between CPU and GPU.
Any CUDA IPC based asynchronous IO support will also benefit \OSPD.

\paragraphb{\textbf{Per-user CVM Deployment without Consensus}}
In \S\ref{sec:threat-model}, we assume all users and the LLM provider trust the shared software stack.
This consensus may not be practical when considering the OS has a large attack surface.
\OSPD is portable to per-user CVM instead of per-user process deployment.
This deployment does not require the consensus on software stack, as each user can independently trust their own stack.
However, to secure model confidentiality, this setup requires the cloud provider to restrict outbound network from per-user CVMs.
This means that the LLM provider must trust the cloud provider, or both being the same party, e.g., Google Gemini and Google Cloud.

\paragraphb{\textbf{Achieve Consensus with A Smart Contract}}
It is interesting to view the initialization of \OSPD's CVM (\S\ref{sec:overview}) from a decentralized perspective.
We do not care which party initializes the CVM, as long as all parties agree on its initial state.
We can standardize the properties and initialization steps of the CVM with a smart contract, hardcoding the target platform, versions of the open source software, minimum number of participants that triggers the initialization and so on.
This smart contract can be certified by an authoritative auditor like CertiK~\cite{certik_smart_contract_audit}, which eases the process of earning trust from users and the LLM provider.

\paragraphb{\textbf{The More Trust, the Better Performance}}
In \S\ref{sec:threat-model}, we assume that all users untrust each other and assign each user a dedicated process.
In practice, some users may trust each other to some extent, e.g., a group of employees in a company.
In this case, \OSPD can assign a process shared by multiple users, reducing overhead of process management and context switching.

\subsection{Limitations and Future Work}
\label{sec:discussion-limitation}

\noindent
\OSPD has its limitations, which bring new opportunities for future works.
We hope our design and discussion will spark further exploration on \name.

\paragraphb{\textbf{Protection of LLM Response}}
\OSPD secures user prompts but not the responses against an untrusted LLM provider.
This implies that the full isolation approaches (\autoref{fig:image2} and \autoref{fig:image3}) are still needed when user would like to secure both the prompts and the responses.
\OSPD offers an alternative approach for different scenarios instead of replacing existing confidential inferencing solutions.

\paragraphb{\textbf{Enhanced TCB}}
As discussed in \S\ref{sec:discussion-mitigating-attacks}, \OSPD inherits the vulnerabilities of the underlying TCB.
Incorporating techniques that enhance the security of \OSPD's TCB, e.g., \cite{duy2025incognitos, qin2023protecting}, into \OSPD can further enhance \OSPD's security against attacks that are out of the current scope.

\section{Related Work}
\label{sec:relwork}

\noindent
In recent years, researchers have explored various approaches beyond confidential computing to preserve user privacy in LLM inference under the assumption of an untrusted LLM provider~\cite{edemacu2024privacy}.

\emph{Differential Privacy} (DP) protects prompt confidentiality by injecting noise into token distributions~\cite{wu2023privacy, panda2023differentially}, generating few-shot random examples~\cite{tang2023privacy}, or tuning the input prompts~\cite{hong2024dp}.
However, these methods are task-specific and compromise \outinv.

\emph{Multi-Party Computation} (MPC)-based methods utilize \emph{secret sharing} that cryptographically splits a number, either an LLM weight or a prompt token, into multiple numbers.
Then they distribute each split to an untrusted party.
The user derives the LLM responses by cryptographically combining the outputs of these parties.
This technique suffers from multiple problems.
First, the untrusted parties must not collude.
Second, secret sharing is not efficient for all LLM operations.
Recognizing its inefficiency, the authors modify the model, e.g., using ReLU instead of SoftMax~\cite{akimoto2023privformer}, or use a much smaller model distilled specially~\cite{li2022mpcformer}, requiring model re-training and violating \outinv.

\emph{Homomorphic encryption} (HE) enables computation on encrypted data and is often combined with MPC to secure user privacy in LLM inference~\cite{liu2023llms, pang2024bolt, hao2022iron, chen2022x}.
However, its significant overhead impedes its use in real-world applications, particularly for nonlinear functions, even in the cases of equipping dedicated hardware.
Recent works~\cite{liu2023llms, pang2024bolt, chen2022x} replace these functions with approximations, which may reduce model accuracy or require model re-training, thereby impeding the use of existing well-trained models.

\emph{Data anonymization} refers to techniques that remove or obscure personally identifiable information (PII) from data to prevent the identification of individuals.
Recent work~\cite{shen2024fire, zeng2024privacyrestore, chen2023hide, kan2023protecting} proposes masking or replacing sensitive segments in prompts, such as names and locations.
However, the anonymization process either fails to protect the secrets or leads to meaningless responses. This occurs when the secrets are essential for the task.
For example, considering a user asks for directions to a specific location, anonymizing the location address will result in an unusable response.

\emph{Obfuscation} generates redundant instances, such as privacy-preserving representations~\cite{yao2024privacy}, pseudo prompts~\cite{mai2023confusionprompt}, and noise tokens~\cite{zhang2024latticegen}, which are mixed with authentic ones to confuse attackers.
The key idea is that attackers cannot distinguish authentic instances from fake ones, whereas users with private prior knowledge can identify the authentic data.
However, these obfuscation based methods usually lead to high computational overhead due to the redundant instances, and are vulnerable to attacks based on statistical analysis.

\section{Concluding Remarks}
\label{sec:conclusion}

\noindent 
Cloud-hosted LLM service is becoming pervasive in our daily lives.
However, it raises privacy concerns since users must submit their prompts to the cloud, which are handled by the LLM service in plaintext.
\OSPD combines confidential computing and \spd (SPD) to protect user prompts from adversaries in the cloud, including both the cloud provider and the LLM provider.
It fully utilizes the confidential computing capabilities of modern hardware to establish trust and protect both user prompts and the LLM.
SPD further secures user prompts from the LLM provider while retaining the full utility of the LLM service, achieving efficient and scalable \name.
Our proposed solution has the potential to enable privacy-preserving LLM applications such as chatbots and AI assistants that involve sensitive data such as personal information, clinical records, and financial documents.

\clearpage
\bibliographystyle{IEEEtran}
\bibliography{IEEEabrv,main}

\clearpage
\appendix
\section{Appendix}

\subsection{Proof of \autoref{thm:partitioned-attention}}
\label{sec:proof_spd}

Let \( Q \in \mathbb{R}^d \) be the query vector. Partition the key and value matrices \( K, V \in \mathbb{R}^{len \times d} \) into \private and \public components:
\[
K = \begin{bmatrix} K_\pvt \\ K_\pub \end{bmatrix}, \quad
V = \begin{bmatrix} V_\pvt \\ V_\pub \end{bmatrix}.
\]
Compute the attention scores \( s \) by:
\[
s = Q K^\top = \begin{bmatrix} Q K_\pvt^\top & Q K_\pub^\top \end{bmatrix}
= \begin{bmatrix} s_\pvt & s_\pub \end{bmatrix},
\]
where \( s_\pvt = Q K_\pvt^\top \) and \( s_\pub = Q K_\pub^\top \).
Define the softmax denominators:
\[
\gamma = \sum_{i=1}^{len} \exp(s_i) = \gamma_\pvt + \gamma_\pub,
\]
with
\[
\gamma_\pvt = \sum_{i=1}^{len_\pvt} \exp(s_{\pvt, i}), \quad
\gamma_\pub = \sum_{i=1}^{len_\pub} \exp(s_{\pub, i}).
\]
The attention output is:
\begin{align}
\softmax(s)V &= \sum_{i=1}^{len} \frac{\exp(s_i)}{\gamma} V_i \nonumber \\
&= \frac{1}{\gamma} \left( \sum_{i=1}^{len_\pvt} \exp(s_{\pvt, i}) V_{\pvt, i}
+ \sum_{i=1}^{len_\pub} \exp(s_{\pub, i}) V_{\pub, i} \right) \nonumber \\
&= \frac{\gamma_\pvt}{\gamma} \left( \frac{1}{\gamma_\pvt} \sum_{i=1}^{len_\pvt} \exp(s_{\pvt, i}) V_{\pvt, i} \right) \nonumber \\
&\quad + \frac{\gamma_\pub}{\gamma} \left( \frac{1}{\gamma_\pub} \sum_{i=1}^{len_\pub} \exp(s_{\pub, i}) V_{\pub, i} \right) \nonumber \\
&= \frac{\gamma_\pvt}{\gamma} \left( \softmax(s_\pvt)^\top V_\pvt \right)
+ \frac{\gamma_\pub}{\gamma} \left( \softmax(s_\pub)^\top V_\pub \right).
\end{align}
Thus,
\begin{align}
\softmax(Q K^\top) V &= \frac{\gamma_\pvt}{\gamma_\pvt + \gamma_\pub} \, \softmax(Q K_\pvt^\top) V_\pvt \nonumber \\
&\quad + \frac{\gamma_\pub}{\gamma_\pvt + \gamma_\pub} \, \softmax(Q K_\pub^\top) V_\pub,
\end{align}
which completes the proof.

\end{document}